\documentclass[prb,aps,epsf,showpacs,preprintnumbers,amsmath,amssymb]{revtex4}
\usepackage{graphicx}% Include figure files
\usepackage{dcolumn}% Align table columns on decimal point
\usepackage{bm}% bold math

\begin{document}

\preprint{Sept.2003}

\title{ Slow, Steady-State Transport with ``Loading'' and Bulk Reactions: 
the Mixed Ionic Conductor La$_2$CuO$_{4+\delta}$ } 
\author{ Wayne M. Saslow} 
\email{wsaslow@tamu.edu}
\affiliation{ Department of Physics, Texas A\&M University, College Station, 
TX 77843-4242}
\date{\today}

\begin{abstract}

We consider slow, steady transport for the normal state of the superconductor 
La$_2$CuO$_{4+\delta}$ in a one-dimensional geometry, with surface fluxes
sufficiently general to permit oxygen to be driven into the sample (``loaded'') either by 
electrochemical means or by high oxygen partial pressure.  We include the bulk
reaction O$\rightarrow$O$^{2-}+2h$, where neutral atoms ($a$) go into ions ($i$)
and holes ($h$). This system is a mixed ionic electronic conductor (MIEC). For 
slow, steady transport, the transport equations simplify because the bulk reaction rate 
density $r$ and the bulk loading rates $\partial_t n$ then are uniform in space and time.  
All three fluxes $j$ must be specified at each surface, which for a uniform current density
$J$ corresponds to five independent fluxes.  These fluxes generate two types of
static modes at each surface and a bulk response with a voltage profile that 
varies quadratically in space, characterized by $J$ and the total oxygen flux $j_O$ 
(neutral plus ion) at each surface. One type of surface mode is associated with 
electrical screening; the other type is associated both with diffusion and drift, and 
with chemical reaction (the {\it diffusion-reaction mode}). The diffusion-reaction mode 
is accompanied by changes in the chemical potentials $\mu$, and by reactions and 
fluxes, but it neither carries current ($J=0$) nor loads the system chemically ($j_O=0$). 
Generation of the diffusion-reaction mode may explain the phenomenon of 
``turbulence in the voltage'' often observed near the electrodes of MIEC's. Within the 
bulk, the local fluxes satisfy a relation that is independent of the applied fluxes. As a 
consequence, the bulk response alone cannot match arbitrary values for the five 
independent input fluxes;  matching occurs by generating appropriate amounts of 
the diffusion-reaction mode at each surface.  The bulk response is completely 
responsible for steady-state loading and typically possesses a voltage profile that 
varies quadratically in space, as for the lead-acid cell. Seven macroscopic 
parameters (three $\partial\mu/\partial n$'s, three diffusion 
constants, and a reaction rate constant) characterize the theory.

\end{abstract}


%72.20.-i, 72.20 Jv, 72.40.+w, 73.50.Pz, , 61.72.Ww, 61.72.Ss ????}

\maketitle

\section {Introduction} 

Mixed ionic electronic conductors (MIECs) are often ``loaded'' with a
specific atom, in either neutral or ionic form.  In La$_2$CuO$_{4+\delta}$, atomic or
ionic oxygen is loaded into a sample in order to produce a concentration that is
more favorable to superconductivity.\cite{AJJ,Kastner}  An idealization of such
an experiment would consider a rod of the material in a one-dimensional geometry
where fluxes of neutral atoms $a$ (O), ions $i$ (O$^{2-}$), and holes $h$ slowly
and steadily enter or leave each end (at $x=0$ and $x={\cal L}$).  Thus, in
addition to the material parameters of the system (three thermodynamic, three
transport, and one reaction), steady state transport requires specification of
six fluxes, subject to the condition that the same net current density $J$
enters and leaves, making a total of five independent fluxes at the
two surfaces.  Hence, although the rod's net oxygen content (neutral plus ion)
increases with time, its net electric charge remains constant. Once the neutrals
or ions enter the rod, they may recombine via the reaction O$\rightarrow$O$^{2-}+2h$.

\subsection{Chemical Loading and Chemical Reactions}

Slow steady transport {\it with loading but without chemical reactions} has already
been studied for the multiple charge-carrier systems of lead-acid
cell\cite{sprl} and La$_2$CuO$_{4+\delta}$.\cite{Saslow1} Because the diffusivities
of the charge-carriers differ, the voltage profile within the bulk has a
component that varies quadratically in space, the details depending on the
fluxes at the surface.  These systems support surface modes of only one type,
which correspond to electrical screening.

Within the context of semiconductors, slow steady transport {\it with chemical 
reactions (electron-hole recombination) but without loading} has also been
studied.\cite{ks}  Because of recombination (a form of chemical reaction) there
is an additional surface mode, but no quadratic variation in space of the voltage profile. 
The additional mode is associated with diffusion and drift,
as well as with recombination, and varies exponentially in space.  Although its
amplitude is dependent on these fluxes, its structure is independent of the
fluxes at the surface.

The present work considers slow steady transport {\it with both loading and chemical
reactions}.  The example again is La$_2$CuO$_{4+\delta}$, since whatever the
excess oxygen content $\delta$, there is a way to force oxygen to enter such
that reactions must take place to cause equilibration.  For example, if
all the oxygen enters the bulk as ions (which can be detected by the
conductivity of the sample\cite{AJJ}), then by employing a high pressure
atmosphere of molecular oxygen one can expect atomic oxygen to enter the sample,
and then convert by reaction to ionic oxygen.  Typically, when oxygen is added
the fraction that goes into atomic bulk states is neither zero nor unity, so
that bulk reactions are needed to cause bulk equilibration.\cite{discuss} 
Therefore, we consider an experiment where the surface fluxes of atomic and ionic
oxygen are completely arbitrary.

In practice, the specific values of the surface fluxes may or may not be known
in a given experiment.  Assuming that they are known, in this work we show how
one may obtain the response of the MIEC, thus giving the rate of loading of O
and O$^{2-}$, the potential, the densities, and the fluxes throughout the sample. 
Our analysis is valid in the limit of low fluxes, where the equations can be
linearized, and for slow steady fluxes.  In principle, the fluxes can vary
slowly in time.  ({\it Slow} is loosely defined relative to time it takes for
the component with the smallest diffusivity to diffuse across the sample.)
Corresponding to the six fluxes specified at the boundaries there must be: at a
given surface, a combination of fluxes that specify the amplitude of the
exponentially-varying {\it diffusion-reaction} mode at that surface; for the
bulk, four independent flux combinations that specify the bulk response, which
gives a uniform chemical loading within the sample and a uniform current flow
through the sample.  To our knowledge, this is the first work on any system to
consider diffusion and drift, bulk reactions, and surface fluxes general enough 
to include chemical loading.

\subsection{``Turbulence'' at the Interface}

Voltage measurement in the area of MIECS is not trivial.  The review by Kudo and
Fueki,\cite{KudoFueki} notes that ``In actual measurements, however, there is
often turbulence of the potential distribution in the vicinity of the electrodes
with which the electric field is applied.''  The authors do not define
``turbulence'' in any operational sense; it may refer to a complex variation in
space or in time.  It has become conventional for experiments in this area to
place the reference electrodes away from the contact between the MIEC and the
adjacent conductors (electronic or ionic), presumably to avoid such
``turbulence.''  See Figure~1. 

\begin{figure}
\includegraphics[width=5in]{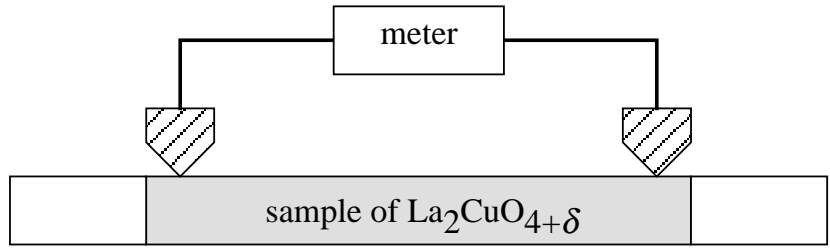}% Here is how to import EPS art
\caption{\label{fig:epsart} Rod of La$_2$CuO$_{4+\delta}$ with two reference 
electrodes attached near its ends.}
\end{figure}

If turbulence refers only to a complex variation in space, then it could be due
to surface modes generated near the interface.\cite{ks}  There are at least two
such modes in systems of this sort: the usual Debye-like screening mode and the
above-described diffusion-reaction mode (in more complex systems, there can be
multiple diffusion-reaction modes).  The reference electrodes measure
``voltages'' $\phi_{ref}$ proportional to electrochemical potentials $\tilde\mu$
($\phi_h=\tilde\mu_h/e$ for hole conduction, and $\phi_i=-\tilde\mu_i/2e$
for divalent anion conduction).  Corresponding to the ``voltages'' are the
``fields'' $E_{h,i}=-\partial_x\phi_{h,i}$.   Since the screening mode involves
no change in the electrochemical potentials, the reference electrodes do not
respond to the screening mode.  On the other hand, the electrodes do respond
to the diffusion-reaction mode.

\subsection{Outline of Paper}

Sect.II employs the methods of irreversible 
thermodynamics, including the effects of chemical reactions, to obtain the 
transport equations for mixed ionics, including the effects of reactions. Sect.III
obtains the steady-state surface modes.  Sect.IV obtains the steady-state bulk
response. Sect.V discusses how to extract the parameters of the system from
measured quantities.  Sect.VI presents our conclusions.

\section {Irreversible thermodynamics} 

Consider, in one-dimension, a uniform sample of La$_2$CuO$_{4+\delta}$, as in Fig.1. 
We take the carriers to be holes and ions (O$^{2-}$) produced by neutral oxygen
atoms O via the reaction
\begin{equation}
\hbox{O}\rightarrow \hbox{O}^{2-} + 2h 
\label{0}
\end{equation}
with reaction rate density $r$. 

\subsection{General Considerations}

Let $u$ denote the energy density, $T$ the
temperature, $s$ the entropy density, $\tilde\mu_h$, $\tilde\mu_i$, and
$\tilde\mu_a$ the hole $h$, ion $i$, and neutral atom $a$ electrochemical
potentials, and $n_h$, $n_i$, and $n_a$ the corresponding number densities. 
Then the fundamental thermodynamic differential for this system is
\begin{equation}
du=Tds+\tilde\mu_h dn_h+\tilde\mu_i dn_i+\tilde\mu_a dn_a.
\label{1}
\end{equation}
With $\mu_h$, $\mu_i$, and $\mu_a$ the chemical potentials, and $\phi$ the 
electrical potential, we have 
\begin{equation}
\tilde\mu_h\equiv\mu_h+e\phi, \quad \tilde\mu_i\equiv\mu_i-2e\phi,   
\quad \tilde\mu_a\equiv\mu_a.
\label{2}
\end{equation}
Here $\phi$ satisfies Poisson's equation
\begin{equation}
\nabla^2\phi=-\frac{\rho}{\varepsilon}
=-\frac{e}{\varepsilon}(-2n_i + n_h),
\label{3}
\end{equation}
where $\varepsilon$ is the dielectric constant, a multiple of the permittivity
of free space $\varepsilon_0$, and the charge density
\begin{equation}
\rho=e(-2n_i + n_h).
\label{rho}
\end{equation}

The conservation laws for this system are, in one-dimension ($x$)
\begin{equation}
\partial_t u+\partial_x j^u=0, \quad
\partial_t s+\partial_x j^s=\frac{\cal P}{ T}\ge 0, 
\label{4}
\end{equation}
\begin{equation}
\partial_t n_h+\partial_x j^h=2r, \quad \partial_t n_i+\partial_x j^i=r,  
\quad \partial_t n_a+\partial_x j^a=-r. 
\label{5}
\end{equation}
Here $j^u$ is the energy flux density, $j^s$ is the entropy flux density, ${\cal
P}$ is the rate density of heat production (${\cal P}/T$ is rate density of
entropy production), $j^h$ is the hole number flux density, $j^i$ is the ion
number flux density, and $j^a$ is the neutral atom number flux density.  In
(\ref{5}), the signs of $r$ are taken such that if O$\rightarrow$O$^{2-} + 2h$,
then $n_h/2$ and $n_i$ increase at the same rate that $n_a$ decreases. The
fluxes, $r$, and $\cal P$ are to be determined.

The time-derivatives of (\ref{1}) and (\ref{4})--(\ref{5}) lead to 
\begin{eqnarray}
0\le {\cal P}&=&-\partial_i j^u +T\partial_x j^s
-\tilde\mu_h(2r-\partial_x j^h)-\tilde\mu_i(r-\partial_x j^i)-\tilde\mu_a(-r-\partial_x j^a)
\nonumber\\
&=&-\partial_x(j^u - Tj^s - \tilde\mu_h j^h - \tilde\mu_i j^i -\mu_a j^a)
\nonumber\\
& &-j^s\partial_x T - j^h\partial_x\tilde\mu_h - j^i\partial_x\tilde\mu_i
- j^a\partial_x\mu_a -r(2\tilde\mu_h+\tilde\mu_i-\mu_a). 
\label{6}
\end{eqnarray}

Expressing ${\cal P}$ as a non-negative quadratic form uniquely requires that: (a) 
\begin{equation}
j^u=Tj^s+\tilde\mu_h j^h+\tilde\mu_i j^i+\tilde\mu_a j^a; 
\label{6a}
\end{equation}
that (b) 
\begin{equation}
j^s =-\frac{\kappa}{ T}\partial_x T
-\alpha_{sh}\partial_x \tilde\mu_h-\alpha_{si}\partial_x \tilde\mu_i-\alpha_{sa}\partial_x \mu_a, 
\label{7}
\end{equation}
where $\kappa\ge 0$ is the thermal conductivity; that (c) 
\begin{eqnarray}
j^h=-\alpha_{hh}\partial_x\tilde\mu_h-\alpha_{hi}\partial_x\tilde\mu_i
-\alpha_{ha}\partial_x\mu_a-\alpha_{hs}\partial_i T,\nonumber\\
j^i=-\alpha_{ih}\partial_x\tilde\mu_h-\alpha_{ii}\partial_x\tilde\mu_i 
-\alpha_{ia}\partial_x\mu_a-\alpha_{is}\partial_i T,\nonumber\\
j^a=-\alpha_{ah}\partial_x\tilde\mu_h-\alpha_{ai}\partial_x\tilde\mu_i 
-\alpha_{aa}\partial_x\mu_a-\alpha_{as}\partial_i T,\label{8}
\end{eqnarray}
where $\alpha_{hh}\ge 0$, $\alpha_{ii}\ge 0$, and $\alpha_{aa}\ge 0$,  
$(\kappa/T)\alpha_{hh}\ge \alpha^2_{hs}$, 
$(\kappa/T)\alpha_{ii}\ge \alpha^2_{is}$, 
$(\kappa/T)\alpha_{aa}\ge \alpha^2_{as}$, 
and $\alpha_{hh}\alpha_{ii}\ge \alpha^2_{hi}$, etc. (by the Onsager 
symmetry principle\cite{Onsager-LL}, $\alpha_{hi}=\alpha_{ih}$, 
$\alpha_{sh}=\alpha_{hs}$, and $\alpha_{si}=\alpha_{is}$); and that (d)   
\begin{equation}
r=-\lambda(2\tilde\mu_h+\tilde\mu_i-\mu_a)=-\lambda(2\mu_h+\mu_i-\mu_a). 
\label{9}
\end{equation}
where the reaction parameter $\lambda\ge 0$ is related to the hole and ion
lifetimes $\tau_h$ and $\tau_i$. In equilibrium $r=0$, so (\ref{9}) implies that
the change in Gibbs free energy be zero, or
$0=2\tilde\mu_h^{(0)}+\tilde\mu_i^{(0)}-\mu_a^{(0)}$, as expected.  For slow
steady processes, $r$ need not be zero but, as we will see, $r$ will take a
value consistent with the surface fluxes, and $r$ will be uniform in space and
in time.

\subsection{Specific Considerations}

From here on, we neglect any coupling to entropy or temperature, and any 
off-diagonal coupling.  Thus 
$0=\alpha_{is}=\alpha_{hs}=\alpha_{as}=\alpha_{hi}=\alpha_{ha}=\alpha_{ia}$.
Moreover, we employ $\alpha_h\equiv\alpha_{hh}$, $\alpha_a\equiv\alpha_{aa}$,
and $\alpha_i\equiv\alpha_{ii}$.

We write the densities in the form
\begin{equation}
n=n^0+\Delta n,
\label{9a}
\end{equation}
where $n^0$ is the equilibrium value and $\Delta n$ is the deviation from
equilibrium ($\delta n$ will be reserved for terms that vary in space, either in
the surface solution or in part of the bulk response).  We now linearize $r$ as
\begin{equation}
r=-\lambda(2\mu_h+\mu_i-\mu_a)
=-\lambda(2{\partial\mu_h\over\partial n_h}\Delta n_h
+{\partial\mu_i\over\partial n_i}\Delta n_i
-{\partial\mu_a\over\partial n_a}\Delta n_a). \label{11}
\end{equation}
Defining the reaction rates $w$ (with dimensions of inverse time) as 
\begin{equation}
w_h=\lambda{\partial\mu_h\over\partial n_h}, \qquad
w_i=\lambda{\partial\mu_i\over\partial n_i}, \qquad
w_a=\lambda{\partial\mu_a\over\partial n_a}, \label{12}
\end{equation}
(\ref{11}) becomes
\begin{equation}
r=-(2w_h\Delta n_h +w_i\Delta n_i -w_a\Delta n_a). \label{13}
\end{equation}
Because the $w$'s vary as the product of a thermodynamic derivative and the reaction 
parameter, they can be thought of as ``thermo-reaction'' parameters. 

Linearizing the fluxes and neglecting the off-diagonal terms, (\ref{8}) becomes 
\begin{eqnarray}
j^h=-\alpha_h\partial_x\tilde\mu_h=-\alpha_h\partial_x(\mu_h+e\phi)
=-\alpha_h({\partial\mu_h\over\partial n_h}\partial_x n_h+e\partial_x \phi),\\
j^i=-\alpha_i\partial_x\tilde\mu_i=-\alpha_i\partial_x(\mu_i-2e\phi)
=-\alpha_i({\partial\mu_i\over\partial n_i}\partial_x n_i-2e\partial_x \phi),\\
j^a=-\alpha_a\partial_x\tilde\mu_a=-\alpha_a\partial_x\mu_a
=-\alpha_a{\partial\mu_a\over\partial n_a}\partial_x n_a. \label{14}\end{eqnarray}
The charge-carrier conductivities $\sigma$ are related to the $\alpha$'s via
\begin{equation}
\sigma_h=e^2\alpha_h, \qquad \sigma_i=4e^2\alpha_i.  
\label{conductivities}
\end{equation}

In terms of the effective electric fields for the charge carriers, 
\begin{equation}
E_h=-\partial_x\phi_h, \qquad E_i=-\partial_x\phi_i,
\label{effectivefields}
\end{equation}
we have 
\begin{equation}
j^h=\alpha_h E_h, \qquad j^i=-2\alpha_i E_i. 
\label{effectivefluxes}
\end{equation}

The diffusivities $D$ are given by 
\begin{equation}
D_h=\alpha_h{\partial\mu_h\over\partial n_h}, \qquad 
D_i=\alpha_i{\partial\mu_i\over\partial n_i}, \qquad
D_a=\alpha_a{\partial\mu_a\over\partial n_a}. 
\label{diffusivities}
\end{equation}
Because the $D$'s vary as the product of a thermodynamic derivative and a
transport parameter, they can be thought of as ``thermo-transport'' parameters. 
Note that the ratio of a $w$ to a $D$ depends only upon reaction and transport,
the thermodynamic derivatives cancelling.  Such ratios occur for the 
diffusion-reaction surface mode. 
 
For completeness, we observe that the electric current density is given by 
\begin{equation}
J=e(j^h-2j^i), 
\label{J}
\end{equation}
and the net oxygen mass flux is given by 
\begin{equation}
j_O=j^i+j^a. 
\label{j_O}
\end{equation}

\section {Steady-state surface modes} 

For surface modes we will employ the notation $\delta n$ for the deviation of a
number density from equilibrium, and $\delta \phi$ for the deviation of the
potential from equilibrium.  We also assume that there is a local relation 
between $n$ and $\mu$, so that (\ref{2}) can be employed to relate $\delta n$ 
to $\delta\tilde\mu$ and $\delta\phi$.  Then the linearized form of (\ref{3}) 
becomes  
\begin{equation}
\nabla^2\delta\phi={e\over\varepsilon}(2\delta n_i-\delta n_h)
={e^2\over\varepsilon}({\partial n_h\over \partial\mu_h}+
4{\partial n_i\over \partial\mu_i})\delta\phi
+{e\over\varepsilon}(2{\partial n_i\over \partial\mu_i}\delta\tilde\mu_i
-{\partial n_h\over \partial\mu_h}\delta\tilde\mu_h).
\label{44}
\end{equation}

\subsection {Screening Surface Mode} 

A steady-state ($\partial n/\partial t=0$) solution of the equations occurs for
no shift in the electrochemical potentials ($\delta\tilde\mu=0$).  In this case
local equilibrium holds, so there are no fluxes and the recombination rate
$r=0$.  As noted above, this mode cannot be measured by electrodes sensitive to individual
electrochemical potentials,\cite{Hebb} although the associated density changes
might be susceptible to other types of analysis (optical, chemical, or
otherwise).  Setting all the electrochemical potentials to zero in (\ref{44})
then gives
\begin{equation}
\nabla^2\delta\phi={e^2\over\varepsilon}({\partial n_h\over \partial\mu_h}
+4{\partial n_i\over \partial\mu_i})\delta\phi. 
\label{45}
\end{equation}
This has solution  
\begin{equation}
\delta\phi=A\exp(-x/l)+B\exp(x/l), 
\qquad {1\over l^2}={e^2\over\varepsilon}
({\partial n_h\over \partial\mu_h}+4{\partial n_i\over \partial\mu_i}). 
\label{46}
\end{equation}
This {\it screening mode} has {\it screening, or Debye, length} $l$ given by
(\ref{46}).  This mode depends only upon equilibrium properties of the system. 
Note that the neutral atoms do not participate at all.  

\subsection {Diffusion and Reaction Surface Mode} 

Another steady-state ($\partial n/\partial t=0$) solution is obtained by
considering the case where $r\ne 0$. Comparison of the three terms in the
continuity equations (\ref{5}) gives the conditions
\begin{equation}
j^h/2=j^i=-j^a. 
\label{46a}
\end{equation}
Note that (\ref{46a}) implies $J=0$ (no current flow) and $j_O=j^i+j^a=0$ (no net
oxygen flow). By the flux equations (\ref{8}), the variations in electrochemical
potential then satisfy
\begin{equation}
-\alpha_h\delta\tilde\mu_h/2=-\alpha_i\delta\tilde\mu_i=\alpha_a\delta\tilde\mu_a.   
\label{47}
\end{equation}
Thus (\ref{9}) can be rewritten as 
\begin{equation}
r\approx-2\lambda\alpha_h\delta\tilde\mu_h({1\over\alpha_h}+{1\over4\alpha_i}+{1\over4\alpha_a}).
\label{48}
\end{equation}
Using this in the steady-state version of (\ref{5}) for $j^h$ then gives 
\begin{equation}
-\alpha_h\partial_x^2\tilde\mu_h
=-4\lambda\alpha_h\delta\tilde\mu_h({1\over\alpha_h}+{1\over4\alpha_i}+{1\over4\alpha_a}).
\label{49}
\end{equation}
This is solved by 
\begin{equation}
\delta\phi=A\exp(-x/L)+B\exp(x/L), 
\qquad {1\over L^2}=4\lambda({1\over\alpha_h}+{1\over4\alpha_i}+{1\over4\alpha_a}).
\label{50}
\end{equation}
This {\it diffusion-reaction mode} has {\it diffusion-reaction length} $L$ given
by (\ref{50}).  All three mobile species contribute to this mode.  Since the
$\alpha$'s are related to diffusion, this mode depends both upon diffusion
($\alpha$) and reaction ($\lambda$).  The faster the diffusion (i.e. the larger
the $\alpha$), the larger the $L$; the faster the reaction (i.e. the larger the
$\lambda$), the shorter the $L$.  This qualitative dependence on diffusion and
reaction is as expected.  A similar mode occurs for semiconductors, where the 
chemical reaction involves the recombination of electrons and holes.\cite{ks} 

If any of these mobile species has a very slow rate of transport, corresponding
to a small $\alpha$, then $L$ is very short.  As a consequence, within a short
distance of the surface the system can adjust from surface-determined boundary
conditions to the bulk values.

For this mode, if $\delta\tilde\mu_h=(C/\alpha_{h})\exp(-x/L)$, then by (\ref{47}) 
\begin{equation}
\delta\tilde\mu_h={C\over\alpha_h}\exp(-x/L), \qquad 
\delta\tilde\mu_i={C\over2\alpha_i}\exp(-x/L), \qquad 
\delta\mu_a=-{C\over2\alpha_a}\exp(-x/L). 
\label{51}
\end{equation}
By (\ref{44}) the potential $\delta\phi$ is given by 
\begin{equation}
({1\over L^2}-{1\over l^2})\delta\phi
=-{e\over\varepsilon}({\partial n_i\over\partial\mu_i}{1\over\alpha_i}
-{\partial n_h\over\partial\mu_h}{1\over\alpha_h})C\exp(-x/L).
\label{52}
\end{equation}
This equation relates $A$ of (\ref{50}) to $C$ of (\ref{51}).  

Since the potential $\delta\phi$ of (\ref{50}) is non-zero for the diffusion-reaction mode, 
this mode also involves the electric field. 

Finally, note that associated with any steady-state surface mode, where
$\partial_t n=0$, there is no deposition of material near the surface.  For the
present mode, where there are both fluxes and chemical reactions, whatever
component is produced by chemical reactions ($r\ne0$) is taken up by nonuniform 
flux ($\partial_x j\ne0$). 

If $j_O=j^i+j^a$ is non-zero for the bulk response, the diffusion-reaction mode
(with $j_O$ non-zero but $j^i$ and $j^a$ individually non-zero) has the
important property that, when added to the bulk mode, it can change the flux
ratio $j^i/(j^i+j^a)$ on moving from the surface to a few $L$ within the bulk.

\section {Steady Bulk Transport Response} 

There are two types of bulk transport response, according to whether or not the 
system is being chemically loaded.  

\subsection {Steady Bulk Transport Response -- No Chemical Loading} 

One solution of the steady-state equations (where $\partial_t n_h=0$, etc.)
occurs for $\delta n_h=\delta n_i=\delta n_a=0$, but $\delta\phi=-Ex+D$,
corresponding to a uniform shift in the electrical potential and a uniform field
$E_x=E$.  This leads to constant partial currents $j^h=\alpha_h E$,
$j^i=-2\alpha_i E$, $j^a=0$, and corresponds to {\it no} chemical loading,
although there is a net oxygen flux that crosses the system.  The solution we
obtain in the next section is sufficiently general that it includes this case,
which will serve as a check on the results of the next section.

\subsection {Steady-state Bulk Flow with Chemical Loading, so $\partial_t n\ne0$} 

Consider a situation with two electrode surfaces, one at $x=0$ and the other at
$x={\cal L}$.  Let there be slow steady flow, and at the surfaces let the ion,
atom, and hole fluxes $j^i$, $j^a$ and $j^h$ be specified, subject to equal
net electric current densities $J$ at the surfaces (so there is no electrical
charging). However, there can be unequal net oxygen fluxes $j^i+j^a$ at the
surfaces, so a net amount of oxygen can be {\it loaded} into the system (mass
charging).  Thus, we specify the fluxes at the boundaries, and within the system
we must determine the potential, the fluxes, and the densities.  This leads to
five input fluxes, which may be thought of as $J$, the two net oxygen fluxes
$j^i+j^a$ at each surface, and the flux ratios $j^i/(j^i+j^a)$.

As a guide to solving the present problem, note that Ref.~\onlinecite{sprl}
considered, in one dimension, slow steady discharge of a lead-acid cell for
concentrations where reactions were unimportant.  (Two and three dimensions have
also been considered.\cite{yangsas})  It was found that the mass loading (and for
discharge, unloading), given by $\partial_t n$ (for ions H$^+$ and HSO$_4^-$),
is uniform in space and time, so the background densities decreased uniformly
in space and time.  The continuity equation then implied that the fluxes vary
linearly in space and are constant in time.  It was then found that all of the
vector quantities in this problem -- two fluxes, two density gradients, and the
electric field -- vary linearly in space and are constant in time.  The two ion
densities have the same gradients, but they have a constant offset, leading to a
constant non-zero charge density that is proportional to the current flow.  This
does not violate charge conservation because the 
screening modes at the surfaces can take up the excess charge.

In the present case we assume that the bulk response has a reaction rate density
$r$ and $\partial_t n$'s that are uniform both in space and in time.  As a
consequence of (\ref{rho}), the charge density $\rho$ also is uniform in space
($\partial_x \rho=0$) and in time ($\partial_t \rho=0$).  For fixed $J$ and
$j^i+j^a$ at each surface, but unspecified flux ratios $j^i/(j^i+j^a)$, this
leads to a consistent solution of the equations for this system; since the
equations are linear, the solution is unique.  The uniformity
assumption leads to a local constraint on the fluxes, and thus specifies the
flux ratios at the surfaces.  By adding in appropriate amounts of the
diffusion-reaction mode at each surface, the flux ratio of the total system can
be made to correspond to any experimental values for the net flux ratios.

\subsubsection{General Considerations}

The charge continuity equation is given by 
\begin{equation} 
\partial_t \rho+\partial_x J=0. 
\label{charge}
\end{equation}
The assumption that $\partial_t \rho=0$ implies that $J$ is constant in space
and in time, consistent with our taking $J$ to have the same value at each
surface.  Moreover, $\partial_t \rho=0$ implies there is no electrical charging,
so the electric charge and the field $E$ and potential $\phi$ that it produces
should all be constant in time.  The assumption that $\partial_x \rho=0$
implies, if $\rho$ is non-zero, that the field varies linearly in space, and
that the potential varies quadratically in space, results that we will show are
consistent with the other equations.  These properties are shared by the slowly
discharging lead-acid cell and chemically loaded La$_2$CuO$_{4+\delta}$ without
reactions.\cite{sprl,Saslow1}

Adding the second two equations of (\ref{5}) yields what amounts to conservation
of the sum of O$^{2-}$ ions and O atoms:
\begin{equation}
\partial_t (n_i + n_a)+\partial_x (j^i + j^a)=0. 
\label{59}
\end{equation}
By (\ref{59}), since the $\partial_t n$'s are assumed to be constant in space
and time, the net oxygen flux $j_O=j^i + j^a$ varies linearly in space, and can
be determined in the bulk from its values at the two ends of the sample.

From the continuity equations (\ref{5}), the constancy in space and time of $r$
and the $\partial_t n$'s then implies that the slopes of all the fluxes are
independent of space and time.  Hence, in the absence of surface modes, these
fluxes $j^a$, $j^i$ and $j^h$ are given by a linear interpolation between
their values at the boundaries.  We will see that, because the fluxes vary
linearly in space, so do the other vector quantities in the problem: $\partial_x
n_i$, $\partial_x n_e$, and $E=E_x=-\partial_x\phi$.  (We have already argued 
that $E$ is linear in space.) 

Let us rewrite Poisson's equation (\ref{3}) in terms of $E=-\partial_x\phi$. 
This gives Gauss's Law which, in one-dimension, reads
\begin{equation}
\partial_x E={e\over\varepsilon}(-2n_i+n_h). 
\label{Gauss}
\end{equation}
Taking the $x$-derivative of (\ref{Gauss}), and using the flux equations to
eliminate $\partial_x n_i$ and $\partial_x n_h$, gives
\begin{equation}
\partial^2_x E={e^2\over\varepsilon}
(4{\partial n_i\over\partial\mu_i}+{\partial n_h\over\partial\mu_h})E
+{e\over\varepsilon}({\partial n_i\over\partial\mu_i}{2j^i\over\alpha_i}
	-{\partial n_h\over\partial\mu_h}{j^h\over\alpha_h}). 
\label{Gaussder}
\end{equation}
This is solved by assuming that $E$ is linear in $x$, so $\partial^2_x E=0$.  
Then (\ref{Gaussder}) gives $E$ as a linear combination of the $j$'s.  Since 
the $j$'s vary linearly in $x$, then so does $E$, and our assumption that $E$ is
linear in $x$ is consistent.  Explicitly, (\ref{Gaussder}) gives
\begin{equation}
E=-\partial_x\phi
=-{1\over e({\partial n_h\over\partial\mu_h}+4{\partial n_i\over\partial\mu_i})}
\bigl[-{\partial n_h\over\partial\mu_h}{j^h\over\alpha_h}
	+2{\partial n_i\over\partial\mu_i}{j^i\over\alpha_i}\bigr].
\label{54}
\end{equation}
Since, unless the values of $j^i$ and $j^h$ at the surface conspire to give no
linear term in $E$ of (\ref{54}), $E$ typically will vary linearly in space. 
Thus the potential $\phi$ typically will vary quadratically in space.  Moreover,
$\partial_x E$ typically is uniform in space, so by (\ref{Gauss}) the charge
density $\rho=e(-2n_i+n_h)$ typically is uniform in space, consistent with our 
assumption that $\partial_x\rho=0$. 

Application of $\partial_x\rho=0$ to (\ref{rho}) gives 
\begin{equation}
2\partial_x n_i=\partial_x n_h.
\label{53}
\end{equation}
Eqs.(\ref{54}) and (\ref{53}), and the relations (\ref{14}) then give 
\begin{equation}
2\partial_x n_i=\partial_x n_h
=-\bigl({{\partial n_h\over\partial\mu_h}{\partial n_i\over\partial\mu_i}
\over{\partial n_h\over\partial\mu_h}+4{\partial n_i\over\partial\mu_i}}\bigr)
\bigl(2{j^i\over\alpha_i}+4{j^h\over\alpha_h}\bigr). 
\label{55}
\end{equation}
Since the $j$'s are linear in space, so are these two density gradients. 

We have already assumed that $r$ of (\ref{13}) is independent of $x$, or 
$0=\partial_x r$.  Using (\ref{53}) and (\ref{12}) leads to 
\begin{equation}
\partial_x n_a={4w_h+w_i\over w_a}\partial_x n_i
={4{\partial\mu_h\over\partial n_h}+{\partial\mu_i\over\partial n_i}
\over{\partial\mu_a\over\partial n_a}}\partial_x n_i, 
\label{56}
\end{equation}
so all three density gradients are linear in space.  

As a consequence of (\ref{56}) and (\ref{55}), (\ref{14}) yields the condition   
\begin{equation}
j^a=-\alpha_a({\partial\mu_a\over\partial n_a})\partial_x n_a
={\alpha_a\over\alpha_i}j^i + 2{\alpha_a\over\alpha_h}j^h. 
\label{56b}
\end{equation}
Thus, for the bulk response, $j^a$ has a specific dependence on $j^i$ and $j^h$.
If (\ref{56b}) is not satisfied at either boundary, then diffusion-recombination
modes are generated, with amplitudes determined in subsection~IV.C.

\subsubsection{Specific Considerations}

We can now obtain explicit values for the three fluxes, the field, and the three
density gradients.

(a) With a knowledge of the mass flux $j_O=j^i + j^a$ at each end of the sample,
of the current $J=e(-2j^i+j^h)$, and of the condition (\ref{56b}), one can obtain
explicit values for all three $j$'s.  This gives  

\begin{equation}
j^i={ {j_O-2{\alpha_a\over\alpha_h}(J/e)}
\over{1+{\alpha_a\over\alpha_i}+4{\alpha_a\over\alpha_h}} },
\label{j^i}
\end{equation}
\begin{equation}
j^h={ {2j_O+(1+{\alpha_a\over\alpha_i}(J/e))}
\over{1+{\alpha_a\over\alpha_i}+4{\alpha_a\over\alpha_h}} },
\label{j^h}
\end{equation}
\begin{equation}
j^a={ {({\alpha_a\over\alpha_i}+4{\alpha_a\over\alpha_h})j_O+2{\alpha_a\over\alpha_h}(J/e)}
\over{1+{\alpha_a\over\alpha_i}+4{\alpha_a\over\alpha_h}} }.
\label{j^a}
\end{equation}

(b) With a knowledge of the three $j$'s and (\ref{54}) one can obtain an explicit value
for $E$.  It is sufficiently complex and unilluminating that we do not present it. 

(c) With a knowledge of the three $j$'s and $E$, by the flux equations
(\ref{14}) we can obtain explicit values for the three density gradients
$\partial_x n$.  Since the $\partial_x n$'s all are proportional to $j^a$ (as 
given above), we do not present them here.  Like the three $j$'s and $E$, the
$\partial_x n$'s vary linearly in space.  We measure the associated
deviations $\delta n$ from $x=0$, so that
\begin{equation}
\delta n=\int_0^x\partial_x n\, dx.
\label{integrate_n}
\end{equation}

Let us now write each density $n$ as the sum of its equilibrium value $n^{(0)}$
and its deviation $\Delta n$, where the latter consists of three terms: an
offset term $\Delta$ that is constant both in space and in time (and has yet to
be determined), a term $at$ (where $\partial_t n=a$) that is constant in space
but linear in time (this corresponds to chemical loading at a uniform rate, and
has yet to be determined; we must also satisfy $at\ll n^{(0)}$), and a
spatially-varying term $\delta n$ that is constant in time but has a non-zero
spatial variation (that has in principle been determined):
\begin{equation}
n_e=(n_h^0 + \Delta_h) + a_h t + \delta n_h, 
\quad n_i=(n_i^0 + \Delta_i) + a_i t + \delta n_i, 
\quad n_a=(n_a^0 + \Delta_a) + a_a t + \delta n_a.   
\label{60}
\end{equation}
Note that $n_i^0=2n_h^0$, since the system is neutral in equilibrium, and $n_a^0$
is (in principle) known from the equilibrium thermodynamics.

We can now obtain explicit values for the three $a$'s. 

(a) Since we assumed that $\partial_t\rho=0$, by (\ref{rho}) we have
\begin{equation} 
2\partial_t n_i=\partial_t n_h.   
\label{52b}
\end{equation}
Eq.~(\ref{52b}), when applied to (\ref{60}) gives 
\begin{equation}
a_h=2a_i.  
\label{60b}
\end{equation}

(b) The requirement that $r$ of (\ref{13}) satisfy $\partial_t r=0$, when 
combined with (\ref{60b}), leads to
\begin{equation}
a_a={4w_h+w_i\over w_a}a_h
={4{\partial\mu_h\over\partial n_h}+{\partial\mu_i\over\partial n_i}
\over{\partial\mu_a\over\partial n_a}}a_h. 
\label{61}
\end{equation}

(c) Since $j^i$ and $j^a$ are known by linearly interpolating their values at
the boundaries, $j^i+j^a$ is known.  Hence (\ref{59}) gives 
\begin{equation}
a_h+a_a=-\partial_x(j^i+j^a).  
\label{61a}
\end{equation}
This, with (\ref{60b}) and (\ref{61}), then determines all three $a$'s, thus specifying 
the time-dependences of the $n$'s. 

We can now obtain explicit values for the three $\Delta$'s, which correspond to
deviations from true equilibrium values.

(a) Since $E$ is known from (\ref{54}), by (\ref{Gauss}) so are $\partial_x E$
and $\rho=\varepsilon\partial_x E$.  Thus, by (\ref{rho}), 
\begin{equation}
n_h-2n_i=\Delta_i-2\Delta_h=\frac{\rho}{e}=\frac{\varepsilon}{e}\partial_x E
\label{dens}
\end{equation}
is determined, where $\partial_x E$ is obtained from a version of (\ref{54})
with the fluxes replaced by their (constant) gradients.

(b) Eq.~(\ref{5}) for $\partial_t n_h$ can be used to determine $r$ in terms of
$a_h$ and $\partial_x j^h$.  The three continuity equations have already been
used twice, once for charge, and once for oxygen, so that this is their third
and final allowable use.  From $r$ we can obtain $\Delta_a$ by (\ref{13}),
written as
\begin{equation}
r\approx-(2w_h\Delta_h +w_i\Delta_i-w_a\Delta_a). 
\label{61b}
\end{equation}

(c) The arbitrariness in time-origin for the terms $at$ in the $n$'s permits us
to set one of the $\Delta$'s to zero, so we may take
\begin{equation}
\Delta_h=0.  
\label{60a}
\end{equation}
The relations (\ref{dens}), (\ref{61b}), and (\ref{60a}) determine all three  
$\Delta$'s. 

Thus we have obtained the time variation $at$, the spatial variation $\delta n$, and the 
offset $\Delta$ for each $n$.

Note that the reaction rate density $r$, and the $\partial n/\partial t$'s are
independent of position and of time, as assumed.

The steady-state bulk solution we have obtained is very general, having three
free parameters (the current and the net mass fluxes at each surface), but the
most general steady-state bulk solution has five free parameters.  By adding in
a diffusion-reaction mode at each surface, we can maintain the same current and
net mass fluxes at each surface, yet still permit five free parameters.  Hence,
since the equations are linear, in obtaining {\it one} steady-state solution, we 
have obtained {\it the} steady-state solution.

The results above hold even if the fluxes are slowly varying in time, provided
one uses the instantaneous values of the fluxes.  In this case the $a$'s also
are slowly varying with time, but their dominant dependence on time is given by
the above approach.

\subsection{Using Surface Solutions to Match the Boundary Conditions}

Assuming arbitrary fluxes at the boundaries, we now determine how much of each
type of response (bulk and surface) is generated.

First, note that (\ref{56b}) applied to our specific bulk response (to which we
append the subscript $b$), can be rewritten as
\begin{equation}
j_b^a={\alpha_a\over\alpha_i}j_b^i+2{\alpha_a\over\alpha_h}j_b^h. 
\label{62}
\end{equation}
Now note that the electric current density of (\ref{J}), 
\begin{equation}
J=e(j_b^h-2j_b^i), 
\label{63}
\end{equation}
is uniform in space, so it holds both in bulk ($b$) and at edges (E).  With
(\ref{63}), (\ref{62}) can be rewritten as
\begin{equation}
j_b^a=(4{\alpha_a\over\alpha_h}+{\alpha_a\over\alpha_i})j_b^i
+2{\alpha_a\over\alpha_h}{J\over e}.  
\label{64}
\end{equation}
At either edge E, write 
\begin{equation}
j_b^a=j_E^a-\Delta j_E^a, \qquad j_b^i=j_E^i+\Delta j_E^a,  
\label{65}
\end{equation}
where $j_E^a$ and $j_E^i$ are the (in principle known) total O and O$^{2-}$ 
fluxes at the edge, and $\Delta j_E^a=-\Delta j_E^i$ are the amplitudes due to the 
surface solutions.  Substituting both parts of (\ref{65}) into (\ref{64}) then 
yields 
\begin{equation}
(1+{\alpha_a\over\alpha_h}+4{\alpha_a\over\alpha_i})\Delta j_E^a
=j_E^a - ({\alpha_a\over\alpha_h}+4{\alpha_a\over\alpha_i})j_E^i 
- 2{\alpha_a\over\alpha_h}{J\over e}
=j_E^a - {\alpha_a\over\alpha_i}j_E^i - 2{\alpha_a\over\alpha_h}j_E^h. 
\label{66}
\end{equation}
Hence, from a knowledge of all three fluxes at an edge E, and the ratios
${\alpha_a/\alpha_h}$ and ${\alpha_a/\alpha_h}$, the associated surface solution
amplitude $\Delta j_E^a$ can be determined.  By (\ref{46a}), $\Delta
j^h_E/2=\Delta j^i_E=-\Delta j^a_E$.  

Global electroneutrality may be maintained by adding in screening modes, with
the appropriate amplitude, at each surface.  The specific value of the screening
mode amplitudes will depend upon the nature of the contact at the surface,
including the electrical contact resistance.  Moreover, there can be a dipole
layer at the surface, so the La$_2$CuO$_{4+\delta}$ itself need not be globally
electroneutral, only the La$_2$CuO$_{4+\delta}$ and a small region at each end of 
the adjacent materials.

The results of this section are sufficiently general that they include those of 
Sect.IV.A, where there are no density gradients in the bulk, and uniform flow of
ions ($j^i$) and holes ($j^h$).  Note that if the surface values of $j^i$ equal the bulk
values, then no diffusion-reaction modes are generated at the surfaces.  If the
surface values of $j^i$ are zero (i.e. high pressure oxygen on one end and
low-pressure oxygen on the other end), then the input flux of oxygen must come
in the form of atoms with $j^a$ non-zero, and the diffusion-reaction modes then
permit the system to convert atoms to ions within the diffusion-reaction length
of the surface.  

\section{On comparison to Experiment}

Comparison to experiment requires a knowledge of the three thermodynamic
derivatives $\partial\mu/\partial n$, the three diffusivities $D$ (or the three
$\alpha$'s), and the reaction constant $\lambda$, which are independent of the
fluxes applied to the system.  Inclusion of the five independent fluxes at the
surface means that the theory for steady state transport involves twelve
independent quantities.  The question then is how they may be determined. 

In principle, equilibrium measurements can yield the three $\partial\mu/\partial n$'s.
Measurement of $l$ for the screening mode can provide an additional constraint
on the $\partial\mu/\partial n$'s.

The fluxes and fields do not depend upon the reaction parameter $\lambda$.  It
appears that $\lambda$ most easily can be obtained by measuring $L$ for the
diffusion-reaction mode, provided that the $\partial\mu/\partial n$'s already
are known.

Dc current measurements with electrodes that do not permit chemical loading, 
and ac current measurements with any set of electrodes, give $\sigma_i+\sigma_h$.  

Measuring electrodes for holes and ions yield effective fields 
\begin{equation}
E_h={j^h\over\alpha_h}, \qquad E_i=-{j^i\over2\alpha_i}, 
\label{E_eff}
\end{equation}
Thus measurements with either type of blocking electrode (for holes or ions) 
gives two combinations of coefficients, one for $j_O=j^i+j^a$ and one for $J/e$.  
Because the equations for the effective fields hold for all $j_O$ and $J/e$, these 
coefficients are overdetermined.

\section{Conclusions}

We have developed the theory of slow steady transport for La$_2$CuO$_{4+\delta}$,  
including both loading and chemical reactions.  We have employed the principles of
irreversible thermodynamics, in which appear the thermodynamic derivatives
$\partial\mu/\partial n$, the diffusion constants $D=\alpha(\partial\mu/\partial
n)$, and the reaction constant $\lambda$.  There are two surface modes, one
associated with screening, and the other associated both with diffusion and drift,
and with chemical reactions.  The screening mode is an equilibrium response,
with no fluxes at all.  The diffusion-reaction mode has no current flux or mass
flux, but non-zero atom, ion, and hole flux.  For a given uniform current
density $J$ and mass flux at each surface, the system has a special bulk
response where the reaction rate density and the mass loading are uniform in
space and in time.  By adding in the diffusion-reaction modes at each surface,
general steady-state solutions are obtained.

As a consequence of the generality of the boundary conditions considered, the
present work applies both to oxygen partial pressure loading\cite{AJJ} and electrochemical
loading\cite{Kastner} of La$_2$CuO$_{4+\delta}$.  Moreover, given the complex nature
of the bulk solution, it is highly unlikely that the material parameters are
such that either oxygen partial pressure loading or electrochemical loading will be able to
avoid generating diffusion-reaction modes.  By varying the imposed current and
mass fluxes at each surface, it should be possible to obtain a number of
constraints on the parameters appearing in the theory.   Thus it would be of
great interest to be able to apply the present results to an actual system. 
Moreover, the diffusion-reaction may be responsible for the puzzling 
phenomenon of ``turbulence of the voltage'' near the electrodes, and measurement 
of its length $L$ would further constrain the values of the material parameters.  

\begin{acknowledgements}
I would like to thank Allan Jacobson for many valuable conversations.  I also
would like to acknowledge the Department of Energy for their support under DOE
Grant No. DE-FG03-96ER45598.
\end{acknowledgements}

%\begin{thebibliography}{}

%\end{thebibliography}

\begin{references}
\bibitem{discuss} Ref.~1 shows that, for an electrochemically determined 
charge $Q$, La$_2$CuO$_{4+\delta}$ satisfies $\delta=Q/2$ for $\delta$ up about 
0.22.  However, the hole content $p$ approximately satisfies $p=2\delta$ for 
$\delta<0.03$ and $p=1.30\delta+0.019$ for $\delta>0.03$.  It thus appears
that the oxygens completely ionize for $\delta<0.03$, but that they 
incompletely ionize for $\delta>0.03$.  The detailed nature of the partially 
ionized oxygen state is not at issue here; in the present context its properties 
can be described phenomenologically in terms of a thermodynamic derivative.  
Ref.~1 proposes two specific models for incomplete ionization of oxygen.
\end{references}
\end{document}